\documentclass[letter,10pt]{article}
\usepackage{makeidx}
\usepackage{graphicx}
\begin{document}
\title{\textbf{Reduction and possible elimination of coating thermal noise using a rigidly controlled cavity with a QND technique}}
\author{Kentaro Somiya\\
\small{\it{Theoretical Astrophysics, California Institute of Technology, Pasadena, California, 91125}}}
\date{}

\twocolumn[
\maketitle
{\footnotesize Thermal noise of a mirror is one of the most important issues in high precision measurements such as gravitational-wave detection or cold damping experiments. It has been pointed out that thermal noise of a mirror with multi-layer coatings can be reduced by mechanical separation of the layers. In this paper, we introduce a way to further reduce thermal noise by locking the mechanically separated mirrors. The reduction is limited by the standard quantum limit of control noise, but it can be overcome with a quantum-non-demolition technique, which finally raises a possibility of complete elimination of coating thermal noise.\\
\vspace{0.1cm}\\}
\vspace{1cm}
]

\section{Overview}
Brownian motion of the coating layers on a test mass is one of the limiting noise sources for high precision measurements such as gravitational-wave detection~\cite{GW} or cold damping experiments~\cite{cold}. Thermal noise is also the most significant source of the environmental decoherence that prevents us to observe quantum behavior of a macroscopic object~\cite{Entanglement}. Reduction of classical noise is a key for various subjects in the modern Physics.

Several ideas to reduce thermal noise have been proposed~\cite{TNreduction}, among which the mechanical separation of coating layers proposed by Khalili~\cite{Khalili} is the easiest and the most sensational one, especially for an interferometer with high-reflective, multi-layer-coated mirrors. The mechanical separation can be realized by using two mirrors with fewer coatings locked to the anti-resonance so that the reflectivity of the anti-resonant cavity is as high as a single mirror with more coatings. 

\begin{figure}[t]
	\begin{center}
		\includegraphics[width=7cm]{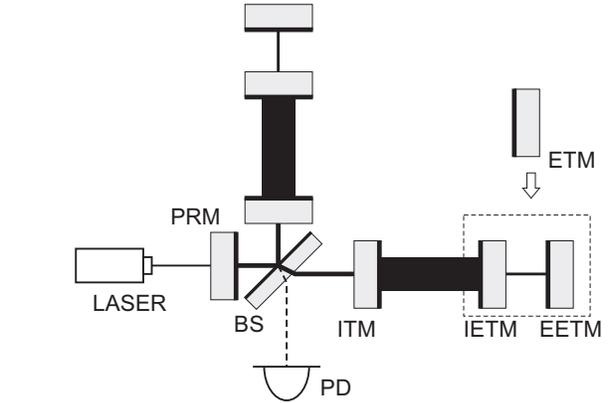}
	\caption{Power-recycled Michelson interferometer with Fabry-Perot cavity in the arms and the end test masses are replaced by an anti-resonant cavity.}
	\label{fig:PRFPMI}
	\end{center}
\end{figure}

Figure~\ref{fig:PRFPMI} shows the configuration of a typical gravitational-wave detector with the end test mass (ETM) replaced by the anti-resonant cavity. The differential mode of the two resonant arm cavities is measured at the dark port of the Michelson interferometer while the common mode and the dc component of the incident light return to the other side, reflected back by the power-recycling mirror (PRM). The reflectivity of the compound mirror in the case without optical losses is given by
\begin{eqnarray}
r_\mathrm{c}=\frac{r_\mathrm{IETM}+r_\mathrm{EETM}}{1+r_\mathrm{IETM}r_\mathrm{EETM}}\ ,
\end{eqnarray}
which is closer to unity than $r_\mathrm{IETM}$ or $r_\mathrm{IETM}$ alone. Coating thermal noise in displacement is proportional to the square root of the coating thickness. The higher the reflectivity of the mirror is, the thicker the coating should be. Thus, the single mirror with higher reflectivity can be replaced by the cavity so that the reflectivity is almost same but coating thermal noise of the input end-test-mass (IETM) is smaller than that of the original ETM. In Ref.~\cite{Khalili}, $r_\mathrm{IETM}$ is assumed to be reasonably high, so that thermal noise of the end end-test-mass (EETM) appears negligible.

The surface of the mirrors toward the inside of the resonant cavity is coated. The light in the resonant cavity probes the motion of the coatings on the IETM, and the light leaking through the anti-resonant end-mirror cavity probes thermorefractive fluctuation of the IETM substrate~\cite{TR} and coating thermal noise of the EETM~\cite{Harry}. The reflectivity balance can be optimized so that the total noise level from both mirrors is minimized. It is important to do the same optimization with the PRM and the input test mass (ITM). Any noise on the PRM does not appear on measuring the differential signal, but noise on the beamsplitter (BS) appears instead. If we replace the PRM into each cavity between the BS and the ITM, the 4-mirror couple cavity is quite symmetric for the two anti-resonant compound mirrors.

\begin{figure}[t]
	\begin{center}
		\includegraphics[width=8cm]{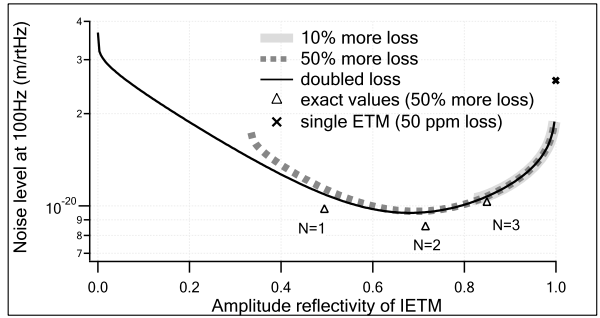}
	\caption{Sum of coating thermal noise from IETM and EETM and thermorefractive noise in IETM at the measurement of the IETM position in terms with the IETM reflectivity. The reflectivity of the EETM is given in such a way that the total optical loss of the end-mirror cavity is 10\%, 50\%, or 100\% more than the loss of a single high-reflective mirror.}
	\label{fig:CoatingBalance}
	\end{center}
\end{figure}

Figure~\ref{fig:CoatingBalance} shows the optimal $r_\mathrm{IETM}$ to reduce the total thermal-noise level. Here $r_\mathrm{EETM}$ is given with each $r_\mathrm{IETM}$ in such a way that the optical loss of the compound mirror is 10\%, 50\%, or 100\% more than a single high-reflective mirror. Here the loss of each mirror is assumed to be 50~ppm. Lowering $r_\mathrm{IETM}$, coating thermal noise of IETM decreases while more light leaks into the end-mirror cavity and probes the motion of EETM. The reflectivity of the mirror coated by the $\mathrm{SiO_2}$-$\mathrm{Ta_2O_5}$ doublets is roughly given by the following equation:
\begin{eqnarray}
r\simeq\sqrt{1-2.8\times0.49^N}\ ,
\end{eqnarray}
with $N$ the number of $\mathrm{Ta_2O_5}$ layers ($\mathrm{SiO_2}$ has one layer fewer). This approximation is fine unless $N<\sim3$; the exact amplitude reflectivity for thin coatings is 0.49 ($N=1$), 0.72 ($N=2$), and 0.85 ($N=3$). Accepting 50\% increase of the optical loss, we obtain the optimal number of the coating layers to be $N=2$. The noise level is improved by a factor of 3 compared with the noise level with a single ETM ($N=15$).

Now, we consider further improvement. Since the optimal result we obtained includes thermal noise from the EETM almost nearly as much as thermal noise from the IETM, the noise level will be even lower if the EETM motion can be completely isolated, or suppressed by the feedback control. In the current gravitational-wave detectors, the mirrors are controlled only at low frequencies so that sensing noise does not impose additional motion to the mirrors~\cite{SomiyaLSC}. Here we use a control field much stronger than the main beam (carrier light) in the end-mirror cavity so that control noise can be sufficiently small. The idea of increasing the control to make a profit has been proposed in previous works~\cite{LRandDOS}, the purpose of which is to obtain gravitational-wave signals by the secondary field that has a better response than the carrier light at some frequencies.

Let us assume that the control field is a frequency-shifted sideband to the carrier light and it resonates in the end-mirror cavity. The sideband senses the motion of the EETM, together with the thermorefractive fluctuation of the IETM, $(1+r_\mathrm{IETM})/(1-r_\mathrm{IETM})$ more than the anti-resonant carrier light. The sideband power being increased, the shot-noise level at the measurement of the end-mirror cavity can be lower than its original thermal fluctuation. On the other hand, quantum radiation pressure noise is imposed on the IETM and it cannot be suppressed by the control. There exists a quantum limit of excess control noise, which indeed can be exceeded by one of the quantum-non-demolition (QND) techniques. We will see this in the following sections with the detailed calculation of quantum radiation pressure noise.

\section{Quantum limit of control noise}

\begin{figure}[htbp]
	\begin{center}
		\includegraphics[width=5cm]{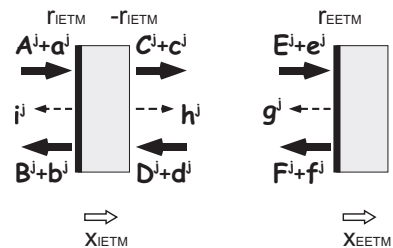}
	\caption{Input -output relation of the end-mirror cavity. Capital letters and small letters indicate classical fields and vacuum fluctuations. Superscript $j$ identifies the carrier (c) or the control sideband (s). Dashed lines are for loss vacuums.}
	\label{fig:inout}
	\end{center}
\end{figure}

\begin{figure}[t]
	\begin{center}
		\includegraphics[width=8cm]{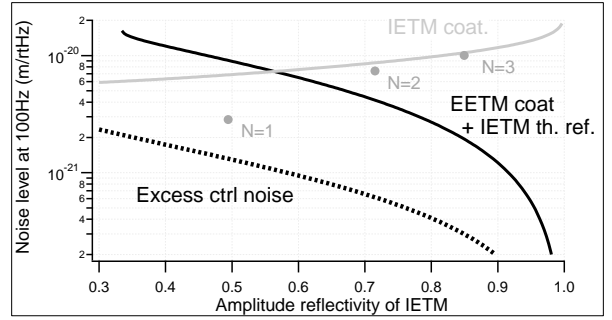}
	\caption{Comparison of excess thermal noise and excess control noise. Gray circles show the exact value of coating thermal noise from the IETM with $N\leq3$. Locking the end-mirror cavity, we can replace thermal noise of the cavity to control noise that is smaller. Thermal noise from the IETM coating appears equally regardless of the control.}
	\label{fig:excessnoise}
	\end{center}
\end{figure}

Figure~\ref{fig:inout} shows the input-output relation of the classical and vacuum fields in the end-mirror cavity. Each bold letter is a vector with the amplitude quadrature and the phase quadrature, which will be depicted by a subscript $1$ and $2$, respectively. The classical fields have amplitude-quadrature components only and the mirror motion appears in the phase-quadrature component. The only difference between the carrier and sideband is the phase shift between the two mirrors. The output fields $\textbf{B}^j$ have the information of $x_\mathrm{EETM}$, so a certain combination of $\textbf{B}^\mathrm{c}$ and $\textbf{B}^\mathrm{s}$ has no thermal noise of the EETM. Let us ignore the optical losses and the transmittance of the EETM for the simplicity; our purpose is to reduce the coating layers of the IETM so that the transmittance of the EETM is low enough. Feeding back the phase-quadrature information of the control sideband to the EETM so that $x_\mathrm{EETM}$ can be cancelled out, we obtain a new output vector:
\begin{eqnarray}
\textbf{z}=\textbf{b}^\mathrm{c}+
\left (
\begin{array}{@{\,}c@{\,}}
0\\
b_2^\mathrm{s}
\end{array}\right )
\times\frac{A_1^\mathrm{c}(1-r)^2}{A_1^\mathrm{s}(1+r)^2}\ ,\label{eq:z}
\end{eqnarray}
the phase quadrature of which is
\begin{eqnarray}
z_2=a_2^\mathrm{c}-\tilde{r}{\cal K}a_1^\mathrm{c}+\frac{\sqrt{2{\cal K}}}{x_\mathrm{SQL}}\tilde{r}x+\frac{A_1^\mathrm{c}(1-r)^2}{A_1^\mathrm{s}(1+r)^2}a_2^\mathrm{s}-\tilde{r}{\cal K}\frac{A_1^\mathrm{s}}{A_1^\mathrm{c}}a_1^\mathrm{s}\ .\nonumber
\end{eqnarray}
\vspace{-0.8cm}
\begin{eqnarray}
\label{eq:z2}
\end{eqnarray}
Here $r=r_\mathrm{IETM}$, $x=x_\mathrm{IETM}$, and
\begin{eqnarray}
\tilde{r}=\frac{4r}{(1+r)^2}\ ,\ 
x_\mathrm{SQL}=\sqrt{\frac{2\hbar}{m\Omega^2}}\ ,\ 
{\cal K}=\frac{8I_0\omega_0}{m\Omega^2c^2}\ ,
\end{eqnarray}
with $I_0$ as the carrier power on the left side of the IETM, $\omega_0$ as the laser angular frequency, $\hbar$ as the Planck constant, $c$ as the light speed, $m$ as the mass of IETM, and $\Omega$ as the measurement angular frequency. While thermal noise of the EETM is suppressed by the control, shot noise moves the EETM and the motion is sensed by the leaking carrier light. Radiation pressure noise of the control sideband moves the IETM and the motion is directly sensed by the carrier light. Taking the square-sum of each vacuum component of the right-hand-side in Eq.~(\ref{eq:z2}), comparing it with the displacement $x$, and choosing the proper $A_1^\mathrm{c}/A_1^\mathrm{s}$, we obtain the quantum-noise level, with excess control noise being minimized at one given measurement frequency, as
\begin{eqnarray}
x_\mathrm{QN}^\mathrm{min}=\frac{x_\mathrm{SQL}}{\sqrt{2{\cal K}}\tilde{r}}\sqrt{1+\tilde{r}^2{\cal K}^2+2\tilde{r}{\cal K}\frac{(1-r)^2}{(1+r)^2}}\ .\label{eq:minQN}
\end{eqnarray}
The last term of Eq.~(\ref{eq:minQN}) is the contribution of excess control noise, which can be rewritten as
\begin{eqnarray}
x_\mathrm{ctrl}^\mathrm{min}=x_\mathrm{SQL}\cdot\frac{1-r}{2\sqrt{r}}\ .
\end{eqnarray}
Figure~\ref{fig:excessnoise} shows the comparison of excess thermal noise, which is namely the sum of thermorefractive noise in the IETM and coating thermal noise from the EETM, and excess control noise that we have just derived. At 100~Hz, the control-noise level is a factor of $6\sim7$ smaller than the other. The optimal number of layers is $N=1$, and the total noise level is $3.1\times10^{-21}~\mathrm{m}/\sqrt{\mathrm{Hz}}$, which is a factor of $\sim2.5$ better than the lowest level without the control.

Note that, even without excess control noise, the minimum value by changing $\cal K$ in Eq.~(\ref{eq:minQN}) does not reach the standard quantum limit $x_\mathrm{SQL}$. This is due to the reduction of the effective mass. Both the carrier and sideband fields sense the position of the IETM, and the effective mass of the end-mirror cavity is $\tilde{r}$ times smaller than the single ETM.

\section{Exceeding the quantum limit}

There exists a way to remove control noise. Using one of the QND techniques, a so-called {\it variational readout} technique~\cite{BAE}, radiation pressure noise can be cancelled out so that control noise can be infinitely small with infinitely high laser power. Instead of feeding back the phase-quadrature information $b_2^\mathrm{s}$, we can feedback $b_\zeta^\mathrm{s}=b_1^\mathrm{s}\sin{\zeta}+b_2^\mathrm{s}\cos{\zeta}$ to the EETM. Equation~(\ref{eq:z}) becomes
\begin{eqnarray}
\textbf{z}^\mathrm{VR}=\textbf{b}^\mathrm{c}+
\left (
\begin{array}{@{\,}c@{\,}}
0\\
b_\zeta^\mathrm{s}
\end{array}\right )
\times\frac{A_1^\mathrm{c}(1-r)^2}{A_1^\mathrm{s}(1+r)^2}\times\frac{1}{\cos{\zeta}}\ ,\label{eq:}
\end{eqnarray}
the phase quadrature of which is
\begin{eqnarray}
&&z_2^\mathrm{VR}=a_2^\mathrm{c}-\tilde{r}{\cal K}a_1^\mathrm{c}+\frac{\sqrt{2{\cal K}}}{x_\mathrm{SQL}}\tilde{r}x\nonumber\\
&&\hspace{1.5cm}+\frac{A_1^\mathrm{c}(1-r)^2}{A_1^\mathrm{s}(1+r)^2}(a_2^\mathrm{s}+a_1^\mathrm{s}\tan{\zeta})-\tilde{r}{\cal K}\frac{A_1^\mathrm{s}}{A_1^\mathrm{c}}a_1^\mathrm{s}\ .\nonumber\\
\end{eqnarray}
Then, choosing the readout quadrature to satisfy
\begin{eqnarray}
\tan{\zeta}=\tilde{r}{\cal K}\times\left[\frac{A_1^\mathrm{s}(1+r)}{A_1^\mathrm{c}(1-r)}\right]^2\ ,
\end{eqnarray}
and increasing the sideband power $A_1^\mathrm{s}$, control noise can be eliminated at one given measurement frequency. Radiation pressure noise of the IETM from $a_1^\mathrm{s}$ is cancelled by driving the EETM with the same amount of vacuum fluctuation. It is only coating noise of a single layer on the IETM that appears at the measurement of this compound mirror's position.

In fact, even the single layer of coating is not necessary. The amplitude reflectivity of an uncoated silica substrate in the vacuum is not zero but $r=0.184$. Using this reflective substrate as the IETM, we finally realize the position measurement without coating thermal noise and excess control noise.

Note that the minimization of control noise can be done at one given frequency as ${\cal K}$ is a frequency-dependent coefficient. Kimble {\it et al} has proposed a so-called {\it filter cavity} to realize the frequency-dependent tuning of the readout quadrature $\zeta$~cite{KLMTV}. Implementation of such a filter should be considered in our case as well.

\section{Summary and discussions}

Reduction of coating thermal noise is the goal of this study. We started from the previous work by Khalili to realize the mechanical separation of the coating layers using an anti-resonant end-mirror cavity. First we pointed out that one can take more advantage of the separation using a 4-mirror coupled cavity with the optimally balanced numbers of coating layers, and demonstrated the optimization. Second, as the main work of this paper, we suggested to lock the end-mirror cavity by control using a sideband field. Thermal noise from one of the mirrors in the cavity is replaced by shot noise imposed by the control and radiation pressure noise of the sideband. Since excess control noise turned out to be smaller than thermal noise, the total noise level was further improved. At last, we introduced a way to remove control noise by the variational readout technique, with which we can choose the reflectivity of the IETM even lower than that of the single-layer coating, namely no coatings, thus no coating thermal noise.

There are two issues to be discussed here. First, the cancellation of thermal noise strongly relies on the assumption that thermal fluctuations sensed by the carrier light and the sideband are the same. The thermal-noise model in Ref.~\cite{Harry} tells us that the fluctuation does not change with the frequency of the light, but it does change with the location of the beam. In practice, the mode and the beam centering can be not perfectly same between the two fields, so this will be one of the issues. Coating thermal noise sensed from one side of the mirror will be hopefully same as that from the other side, according to the model.


The second issue is about other kinds of coating thermal noise. Thermal expansion due to the Brownian motion or the fluctuation of the refraction index causes the fluctuation of the complex reflectivity of a mirror. These are called thermoelastic noise~\cite{Fejer} and thermorefractive noise~\cite{TRc}, respectively, which are not included in the calculation of this paper. Thermoelastic noise decreases by reducing the coating layers as well as Brownian thermal noise, while coating thermorefractive noise is independent from the number of the layers. Brownian thermal noise tends to be larger than thermoelastic or thermorefractive noise, but it decays faster at higher frequencies, so the final optimization should be done taking the frequency dependence into account. Recently, Evans {\it et al} pointed out that thermoelastic noise and thermorefractive noise of the coating would happen to cancel with a certain number of coating layers~\cite{Matt}. Since the cancellation suggested in this paper also relies on the optimization of the number of coatings, another parameter that works independently on Brownian thermal noise and thermoelastic noise will be necessary to realize both optimizations. 

\section*{Acknowledgement}
The author would like to appreciate Prof. Yanbei Chen for valuable discussions. This research is supported by Japan Society for the Promotion of Science (JSPS).

\bibliographystyle{junsrt}
\pagestyle{headings}

\end{document}